\documentclass[preprint,5p,twocolumn]{elsarticle}

\usepackage{amssymb}
\usepackage{amsthm}
\usepackage{amsmath}
\usepackage{color,verbatim}

\journal{Chaos, Solitons, and Fractals}

\usepackage[usenames,dvipsnames]{xcolor}

\begin{document}

\begin{frontmatter}

\title{Link overlap, viability, and mutual percolation in multiplex networks}

\author[1]{Byungjoon Min, Sangchul Lee, Kyu-Min Lee, and K.-I.\ Goh}

\address[1]{Department of Physics and Institute of Basic Science, Korea University, Seoul 136-713, Korea}

\begin{abstract}
Many real-world complex systems are best modeled by multiplex networks. The multiplexity has proved to have broad impact on the system's structure and function. Most theoretical studies on multiplex networks to date, however, have largely ignored the effect of link overlap across layers despite strong empirical evidences for its significance. In this article, we investigate the effect of link overlap in the viability of multiplex networks, both analytically and numerically. 
Distinctive role of overlapping links in viability and mutual connectivity is emphasized and exploited for setting up proper analytic framework.
A rich phase diagram for viability is obtained and greatly diversified patterns of hysteretic behavior in viability are observed in the presence of link overlap.
Mutual percolation with link overlap is revisited as a limit of multiplex viability problem, and controversy between existing results is clarified. The distinctive role of overlapping links is further demonstrated by the different responses of networks under random removals of overlapping and non-overlapping links, respectively, as well as under several removal strategies. Our results show that the link overlap strongly facilitates viability and mutual percolation; at the same time, the presence of link overlap poses challenge in analytical approach to the problem.
\end{abstract}

\begin{keyword}
Multiplex networks, Link overlap, Viability, Mutual percolation, Network robustness
%% keywords here, in the form: keyword \sep keyword
%% PACS codes here, in the form: \PACS code \sep code
%% MSC codes here, in the form: \MSC code \sep code
%% or \MSC[2008] code \sep code (2000 is the default)
\end{keyword}

\end{frontmatter}

%% main text

\section{Introduction}
\label{sec:intro}

Many real-world complex systems ranging from society to critical infrastructure operate through multiple layers of distinct interactions among constituents as well as the interplay between these interaction layers \cite{NoN,Kivela2014,Boccaletti2014}. 
Multiplex network \cite{KMLee2013} is a class of networks introduced to model such systems, in which the same set of nodes (say, individuals in a society) are connected via more than one type of links (say friendships, kinships, co-workerships, {\it etc}). 
Each type of links in multiplex networks forms the network layer. 
The study on multiplex networks is one of the major contemporary topics of network theory \cite{KMLee2012,Brummitt2012,Evol2012,Boolean2012,Gomez2013,Bianconi2013,Kim2013,Min2014a,Min2014b,Baxter2014}, together with closely related topics such as  interacting networks \cite{Leicht2009}, interdependent networks \cite{Buldyrev2010}, and interconnected networks \cite{Radicchi2013},~{\it etc.}

Introduction of distinct layers in multiplex networks necessitates new conceptual as well as computational developments beyond the well-established single-network framework \cite{NewmanBook}, as one has to properly deal with the coupling between layers as well.
One such factor is the pattern of multiplexity, that is, how the layers are coupled structurally.
In real-world multiplex systems, the coupling structure between layers should by no means be random but correlated.
The term correlated multiplexity was introduced to refer to this correlation property \cite{KMLee2012}. 
Correlation properties of the layer coupling have been assessed empirically \cite{Szell2010} in term of the interlayer degree correlation and the link overlap, and indeed both were found to be significant. 
Effect of the interlayer degree correlation has been theoretically investigated for percolation \cite{KMLee2012,KMLee2013} and network robustness \cite{Min2014a} in multiplex networks (see also related studies on the interdependent networks \cite{Parshani2010,Buldyrev2011}).
The multiplex network model based on the coevolution of layers \cite{Kim2013} has been proposed to model the emergence of interlayer degree correlation in evolving multiplex networks.
A more comprehensive  framework for characterization of the correlation properties of multiplex networks has recently been proposed \cite{Nicosia2014}.

Meanwhile, in social network literature the existence and consequences of link overlaps across social network layers have been the central issues of multiplexity studies, proving to play instrumental roles in social structure and dynamics in diverse ways \cite{Minor1983}.\footnote{Indeed, the term multiplexity itself has often (but not universally) been used as the measure of  degree of link overlap in a social network.}
However, most theoretical studies on multiplex networks thus far have focused on the case of randomly coupled sparse layers, thereby the effect of the link overlap across layers has largely been ignored. Only recently two groups of researchers have investigated explicitly the effect of link overlap for the problem of mutual percolation \cite{Hu2013, Cellai2013} using different theoretical approaches and obtained different conclusions about the existence of tricritical point. 

Mutual percolation is a percolation model originally introduced for interdependent networks \cite{Buldyrev2010,Son2012} yet applicable to multiplex networks straightforwardly \cite{Min2014a}. 
A pair of nodes in multiplex network are said to be mutually connected if the pair is connected through each and every layer in the network.
A set of nodes within which every pair is mutually connected form the mutually-connected component, or the mutual component for short. 
If there exists an extensive mutual component, called the giant mutual component, we say that mutual percolation has occurred and the system is mutual-percolated. 
The mutual percolation is a minimal model of cooperative coupling between layers in multiplex systems. 

Recently the so-called viability of multiplex networks has been introduced \cite{Min2014b} (see also closely related weak percolation models proposed in \cite{Baxter2014}).
The viability of multiplex networks is based on the concept of multiple resource demands for proper functioning in multiplex systems. 
For such kind of systems, a node can be viable only when it is supplied with each and every kind of resources successfully through corresponding network layer. 
As we will see in the following sections, the problem of viability of multiplex networks turns out to include the mutual percolation as a particular limit \cite{Min2014b}, thus provides unifying framework for existing different approaches for mutual percolation with link overlap \cite{Hu2013,Cellai2013}, thereby helps clarify how and why the two studies produce different results. 

Thus we study in this article the effect of link overlap on the viability of multiplex networks \cite{Min2014b} with two-fold objectives: 
i) To present generalized theoretical formalism and understanding of the problem of viability of multiplex networks with link overlap; and ii) to clarify the discrepancy of existing results for the mutual percolation with link overlap by way of the limiting case of the multiplex viability problem.

The rest of the article is organized as follows. In Section~\ref{sec:viability}, we introduce the model of viability in multiplex networks and summarize main results from \cite{Min2014b}. Effect of the link overlap on viability is investigated in Section~\ref{sec:viability-overlap}. 
As a limiting case of the viability problem, we revisit the mutual percolation with link overlap and discuss about the two existing approaches~\cite{Hu2013, Cellai2013} in Section~\ref{sec:mutual}. Implications to the network robustness under random link failures in networks with link overlap are investigated in Section~\ref{sec:robustness}.
Finally summary and discussions in Section~\ref{sec:conclusion} will follow.

\section{Viability of multiplex networks}
\label{sec:viability}

The viability of multiplex networks was introduced \cite{Min2014b} based on the concept of multiple resource demands for proper functioning in multiplex systems such as the civil society~\cite{Rinaldi2001} and biological systems~\cite{Ferreira2002}. For example, in the case of society, our life in modern society relies on interrelated infrastructure networks including water supply networks, gas supply networks, and power grid systems \cite{Rinaldi2001}. In such systems, simultaneous connectivities of a node with the nodes producing resources (thereafter called the resource nodes), such as power plants in the power grid and water sources in the water supply networks, through a series of functioning nodes are essential for the proper functioning, {\it i.e.,} to be viable. In this section we briefly summarize the previous work \cite{Min2014b} to outline the viability model and its solution properties, before generalizing it to cases with overlapping links in the following sections.
For more details and additional results regarding the multiplex viability model the reader is referred to \cite{Min2014b}.

\subsection{The multiplex viability model: Definition and algorithms}
Let us consider a network with $n$-multiple layers, where each layer of the multiplex network corresponds to a certain infrastructural network. A given fraction $\rho$ of resource nodes generates and distributes resources essential to be viable. 
A key assumption of the model is that only viable nodes can function properly and transmit resources further to their connected neighbors.
Then, a node is viable only if it can reach, via the viable nodes, to a resource node in each and every layer.
One is interested primarily in what fraction of the nodes would be viable for given network parameters and distribution of resource nodes.
Specifically, we define the viability $V$ to be the fraction of viable nodes in the limit $N \rightarrow \infty$.

To identify clusters of viable nodes (or viable clusters for short) in a given network, two algorithms were proposed \cite{Min2014b}, called the cascade of activations (CA) and deactivations (CD), respectively. How the two algorithms proceed when applied to a small multiplex network of $N=9$ nodes with two layers (without overlapping links) is illustrated in Fig.~1.

\begin{figure}[t]
\begin{center}
\includegraphics[width=.95\columnwidth]{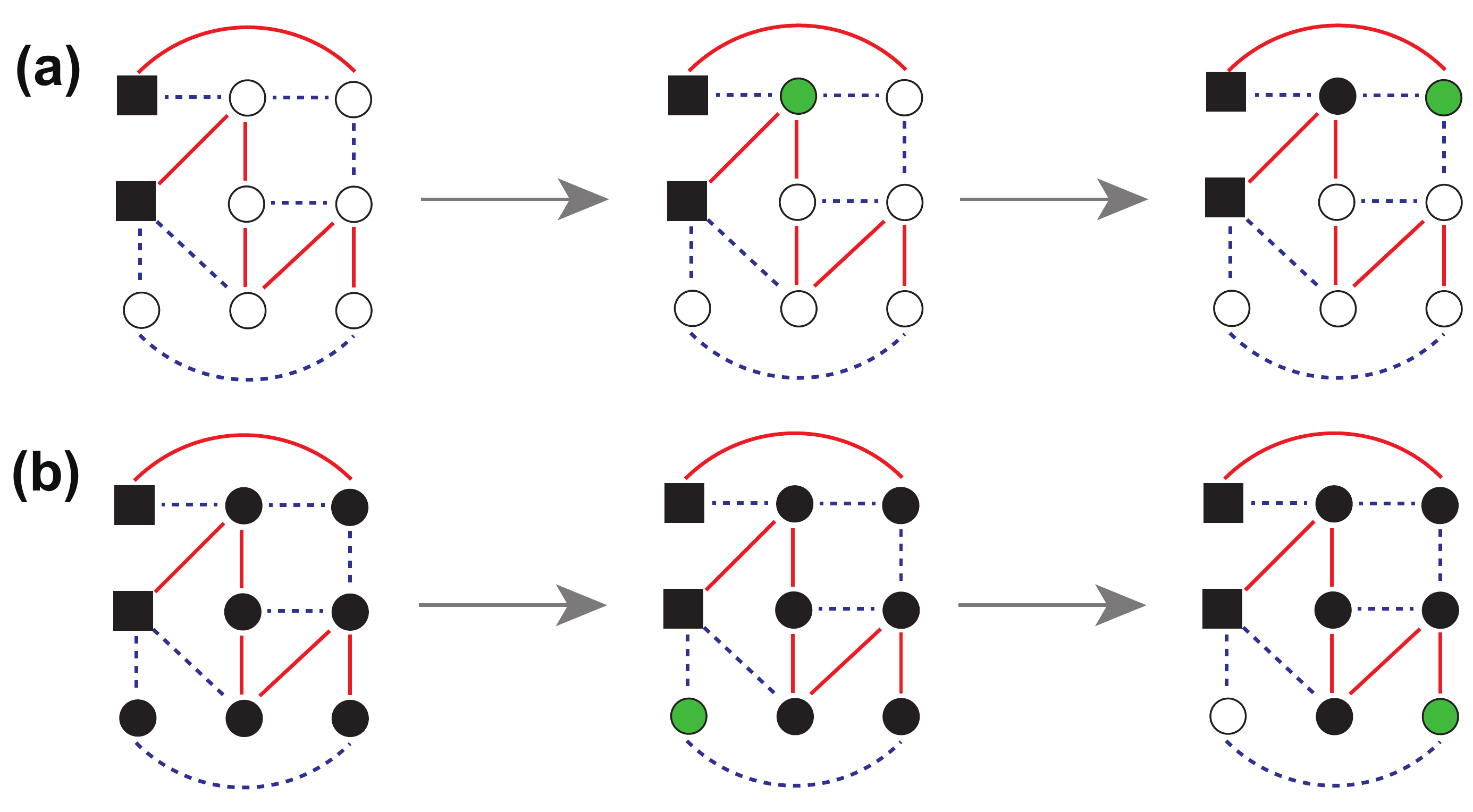}
\caption{(color online)
Illustrations of the iterative algorithms of (a) the cascade of activations (CA) and (b) deactivations (CD).
Source nodes (squares) generate resources. If a node connects with resource nodes 
through each type of links denoted by solid and dashed lines, the node is viable (filled circles) and
can transmit resources further to its neighbors. If not, the node is unviable (open circles).
Shaded (green) circles denote the node whose state is to be updated [activated in (a) and deactivated in (b)] at each step.
Reproduced from Ref.~\cite{Min2014b}. %Courtesy by APS.
}
\end{center}
\label{fig:algorithm-viability}
\end{figure}

\begin{figure*}[t]
\begin{center}
\includegraphics[width=.85\linewidth]{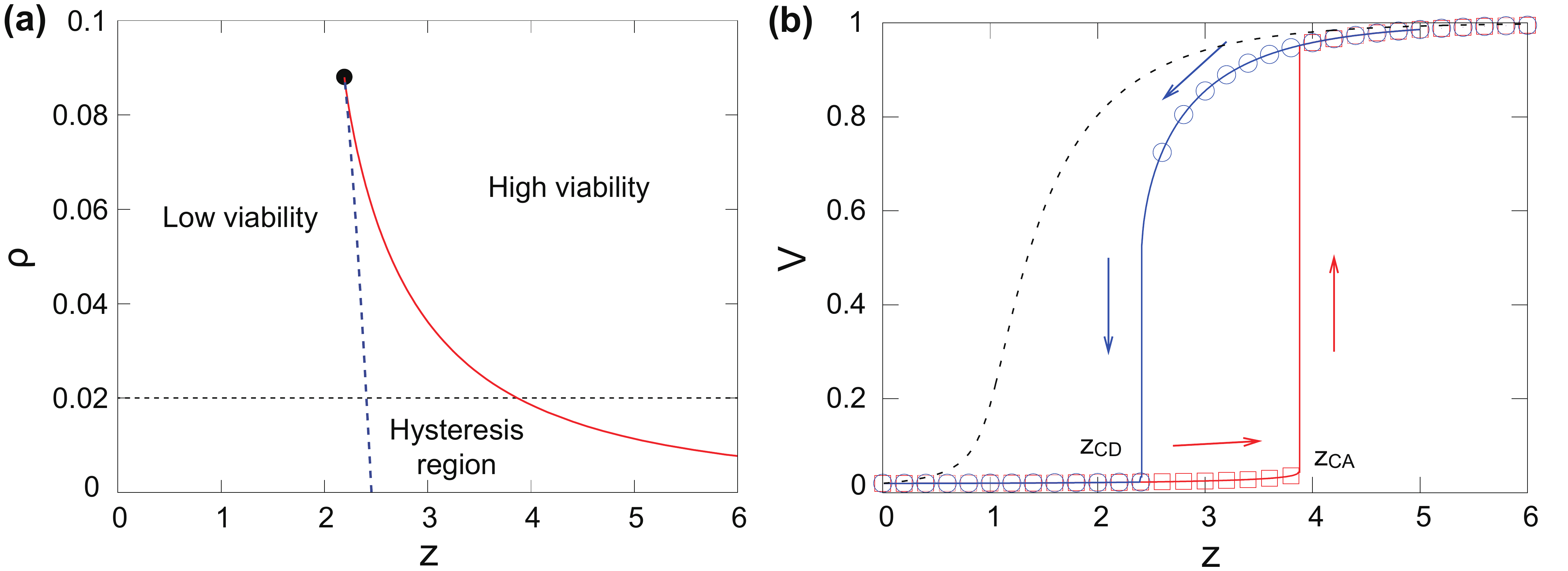}
\caption{
(a) $(\rho, z)$-phase diagram of viability of duplex ER graphs. 
Solid and dashed lines indicate the locations of a discontinuous jump 
for the CA and the CD algorithm, respectively.
Filled circle indicates the critical point at which the discontinuous jumps disappear.
Dotted horizontal line indicates $\rho=0.02$ for which the panel (b) is obtained.
(b) Hysteresis curves with $\rho=0.02$. 
Starting from the well-connected high viability state, the systemic collapse $(\circ)$ and the subsequent recovery $(\Box)$ exhibit different curves. The dashed line indicates the viability for $n = 1$ ER graphs for comparison, which does not exhibit hysteresis.
Adapted from Ref.~\cite{Min2014b}. %Courtesy by APS...
}
\end{center}
\label{fig3}
\end{figure*}

The CA algorithm runs as follows:
\begin{itemize}
\item Initially all nodes except resource nodes are labeled unviable.
\item At each step, unviable nodes that are linked to viable nodes through each and every layer
[colored green in Fig.~1(a)] are activated to become viable. 
\item This cascade of activations continues until no nodes newly become viable.
\end{itemize}
The CD algorithm runs as follows:
\begin{itemize}
\item Initially all nodes including resource nodes are labeled viable.
\item At each step, the nodes that do not reach the resource nodes in any of the layers [colored green in Fig.~1(b)] are deactivated to become unviable.
\item This cascade of deactivations continues until there are no nodes to deactivate.
\end{itemize}

Note that the viable clusters obtained from the CA and CD algorithms can be in general different from each other, as exemplified in Fig.~1.
While the viable nodes for the CA are always viable for the CD as well, it is not always so {\it vice versa}.
This already suggests that the problem of multiplex viability can have multiple solutions, depending on the system's history.

\subsection{Analytic solution for the multiplex viability}

To gain more insight, we compute the viability analytically.
This can be done for the cases of multiplex networks with layers that are locally-tree-like;
we further assume that the layers are coupled randomly, so that the link overlaps among layers are negligible for sparse networks we consider; finally, resource nodes are assumed to be distributed randomly over the network.

In $n$-layer networks, each node has $\vec{k}=(k_1,\dots,k_n)$ degrees over different layers drawn from 
the joint degree distribution $p(\vec{k})$.
To calculate viability, we first consider the probability $u_i$ 
that the node reached by following a randomly chosen $i$-type link ($i=1,\dots,n)$ is not viable.
Given the initial fraction $\rho$ of randomly distributed resource nodes on locally-tree like structures,
$u_i$'s satisfy the following self-consistency equations,
\begin{align}
1- u_i=\rho+(1-\rho)\sum_{\vec{k}} \frac{k_i p(\vec{k})}{z_i} (1-u_i^{k_i-1}) {\prod_{\substack{j=1 \\ j\ne i}}^n} (1-u_j^{k_j}),
\end{align}
where $z_i$ is the mean degree of the $i$-layer network.
The first term is the probability that the chosen node is a resource node.
The second term is the probability that it is 
connected with viable nodes through each type of links.
The viability $V$ is obtained as the probability that a randomly chosen node is viable,
which is given in terms of $u_i$'s by 
\begin{align}
V=\rho+(1-\rho)\sum_{\vec{k}} p(\vec{k}) \prod_i (1-u_i^{k_i}).
\end{align}
By solving Eqs.~(1--2) for given $p(\vec{k})$ and $\rho$, one obtains the viability $V$. 
One may note that when $\rho=0$ Eqs.~(1--2) reduce to those of mutual percolation \cite{Son2012} with the viability $V$ identified with the size of the giant mutual component.

\subsection{Applications to duplex random graphs: Bistability and hysteresis}
We illustrate  basic features of the multiplex viability model with a specific example of the multiplex networks with two randomly-coupled Erd\H{o}s-R\'enyi (ER) layers (duplex ER graphs).
For simplicity the mean degrees of two layers are chosen to be the same, denoted as $z$.
Then Eqs.~(1) and (2) reduce to a single equation for $V$ as
\begin{align}
V=\rho+(1-\rho) (1-e^{-z V})^2.
\end{align}
By solving Eq.~(3), $V$ can be obtained for given $\rho$ and $z$.
Eq.~(3) is found to have two stable solutions for a range of parameters.

One can construct the $(\rho,z)$ phase diagram [Fig.~2(a)], 
in which the low and high viability phases are separated by the two lines of loci of the saddle node bifurcations.
The solid line corresponds to the points at which the viability for the CA algorithm undergoes a discontinuous change, 
and the dashed line indicates that for the CD algorithm.
Between the two lines (hysteresis region), there are two possible stable solutions, and $V$ is determined by its initial value.
For example, for the CA algorithm, $V$ keeps to be in low viable state with increasing $z$ until the abrupt jump at $z_{\text{CA}}$ on the solid line.
For the CD case, high viability sustains until abrupt collapse at $z_{\text{CD}}$ on the dashed line [Fig.~2(b)].
Two lines merge at the critical point located at 
$(\rho_c,z_c)=(\frac {2 \log 2-1}{2 \log 2 +3},\log 2 +\frac{3}{2})$.
Above $\rho_c$, the viability changes gradually without discontinuity as $z$ increases and the distinction between the two phases disappear.

Such a bistability leads to a hysteresis in viability of multiplex networks [Fig.~2(b)].
In order to demonstrate the hysteresis explicitly, 
let us suppose the following scenario of a sequence of systemic collapse 
and subsequent recovery of viability.
Initially the system is in high viability state with well-established networks (that is, sufficiently large $z$).
As $z$ decreases due to random failures of links, $V$ abruptly collapses at $z_{\text{CD}}$.
After collapse, if one tries to restore viability to the level before the collapse,
one needs to increase the link density up to $z_{\text{CA}}$, which is significantly larger than the point of collapse $z_{\text{CD}}$.
This suggests that the multiple resource demands lead to not only a 
potential danger of abrupt collapse but also an excessively severe complication in recovery.
Note that the bistability and hysteresis are absent in the single networks, $n=1$ [Fig.~2(b)].

\section{Impact of link overlap on multiplex viability}
\label{sec:viability-overlap}
In this section, we study the viability of multiplex networks with link overlap. With link overlap, one can still apply the CA and CD algorithms in a straightforward manner (Fig.~3).
However, one should note different roles of overlapping and non-overlapping links for mutual connectivity and viability. 
For ordinary connectivity, making link overlap over different layers does not extend the existing connected clusters but merely provides redundancy.
For mutual connectivity, by contrast, the overlapping links can play distinctive role: 
A cluster formed by overlapping links (hereafter called the overlap cluster for short) is by itself  a mutually-connected component.\footnote{This overlap cluster can be part of a larger mutual component through non-overlapping links.} 
Consequently the whole overlap cluster becomes viable once any of its member nodes becomes viable. Similarly, the overlap cluster becomes deactivated also as a whole. 

In what follows we will focus on the duplex systems. 
Generalizations for $n>2$ cases will be discussed later.
Given the distinct roles of non-overlapping and overlapping links, it is useful to represent the degree 
of a node by distinguishing the two classes. 
Therefore we use the vectorial degree $\vec{k}=(k_1,k_2,k_o)$ for each node where $k_i$ $(i=1, 2)$ is the degree for non-overlapping link of type $i$ and $k_o$ is for overlapping links, all of which are drawn from the joint degree distribution $p(\vec{k})$. We further assume that the layers are randomly coupled 
as well as the overlapping and the non-overlapping degrees are uncorrelated.
We provide analytic treatments for the CA and the CD algorithms, respectively, for viability in the presence of link overlap. 
It is shown that the two algorithms are described by different sets of equations, displaying a rich variety of behaviors in  viability.

\begin{figure}[t]
\begin{center}
\includegraphics[width=.95\columnwidth]{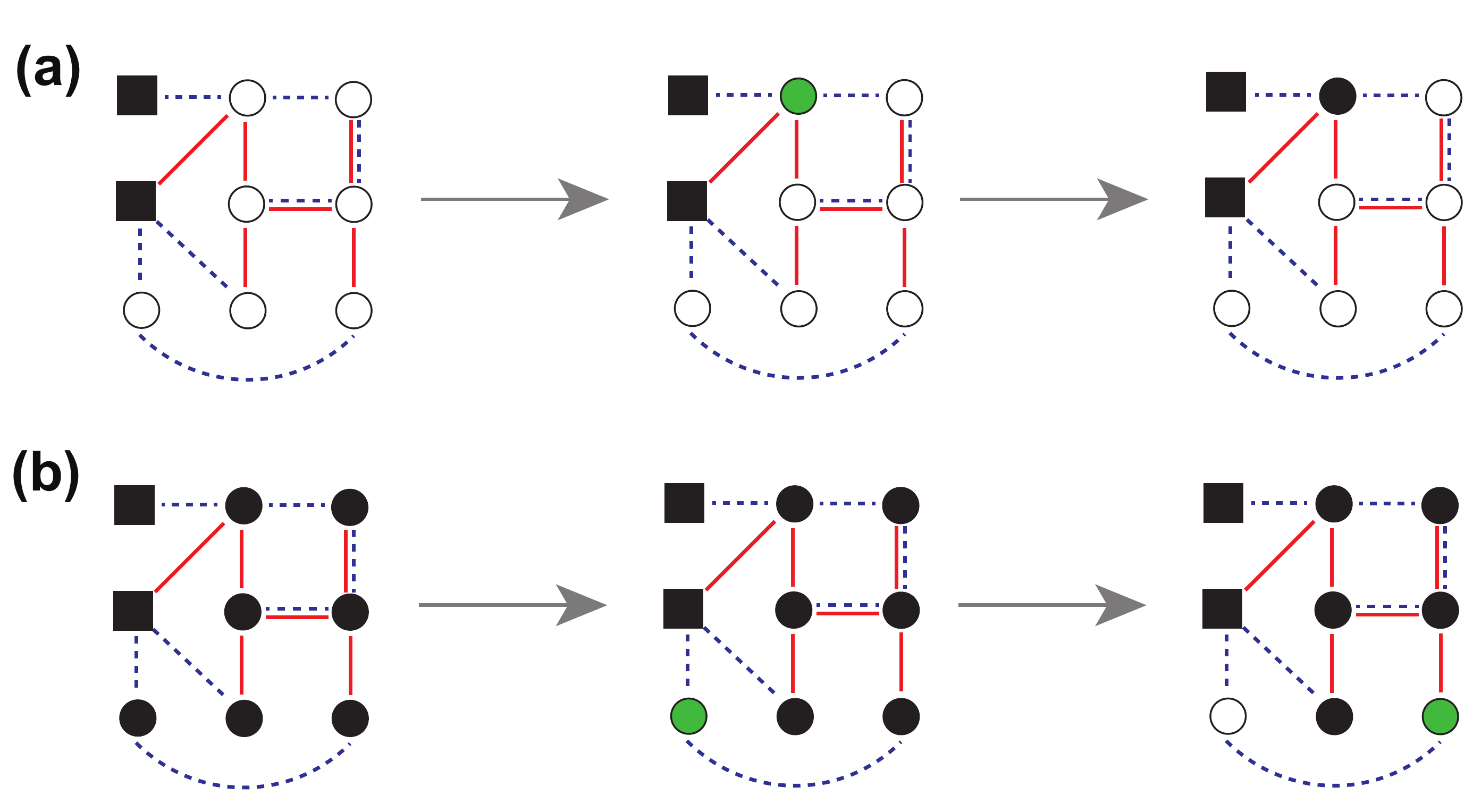}
\caption{(color online)
Illustrations of the iterative algorithms of (a) the cascade of activations (CA) and (b) deactivations (CD) with link overlap. See the caption of Fig.~1 for details.
}
\end{center}
\label{fig:algorithm-overlap}
\end{figure}

\subsection{Analytic solution for CA algorithm with link overlap}
To become viable in the CA algorithm, a node must be connected individually with viable nodes either through both types of links simultaneously or through overlapping links. 
Accounting for this, the self-consistency equations for the probabilities $u_i$ ($u_o$) that the node reached by following a randomly chosen type-$i$ nonoverlapping (overlapping) link is not viable in the CA algorithm thus can be written as
\begin{align}
1-u_i=\rho+(1-\rho) \sum_{\vec{k}} &\frac{k_i p(\vec{k})}{z_i} \Big[ (1-u_o^{k_o})\nonumber\\
&+u_o^{k_o}(1-u_i)^{k_i-1} (1-u_j)^{k_j} \Big], \nonumber \\
1-u_o=\rho+(1-\rho) \sum_{\vec{k}} &\frac{k_o p(\vec{k})}{z_o} \Big[ (1-u_o^{k_o-1})\nonumber\\
&+u_o^{k_o-1}  (1-u_1)^{k_1} (1-u_2)^{k_2} \Big].   
\end{align}
The first term is the probability that the chosen node is a resource node; the second term is the probability that it is connected with viable nodes through overlapping links; and the last term is the probability that it is connected with viable nodes through each type of non-overlapping links but not through overlapping links. Similarly, the final fraction of viable nodes $V$ can be obtained as 
\begin{align}
V=\rho+(1-\rho) \sum_{\vec{k}}  p(\vec{k}) &\Big[ (1-u_o^{k_o}) \nonumber\\ 
&+u_o^{k_o}(1-u_1)^{k_1} (1-u_2)^{k_2} \Big].  
\end{align}

Like Eqs.~(1--2), Eqs.~(4--5) can have two stable solutions for a range of parameters. As illustrated in Fig.~4, the outcome of CA algorithm corresponds to the lower branch solution for $V$ of Eqs.~(4--5).  
In passing, it is worthwhile to note that for $\rho=0$, Eqs.~(4--5) reduce to those for the mutual percolation with link overlap proposed in Ref.~\cite{Cellai2013}.

\begin{figure}[t]
\begin{center}
\includegraphics[width=.95\columnwidth]{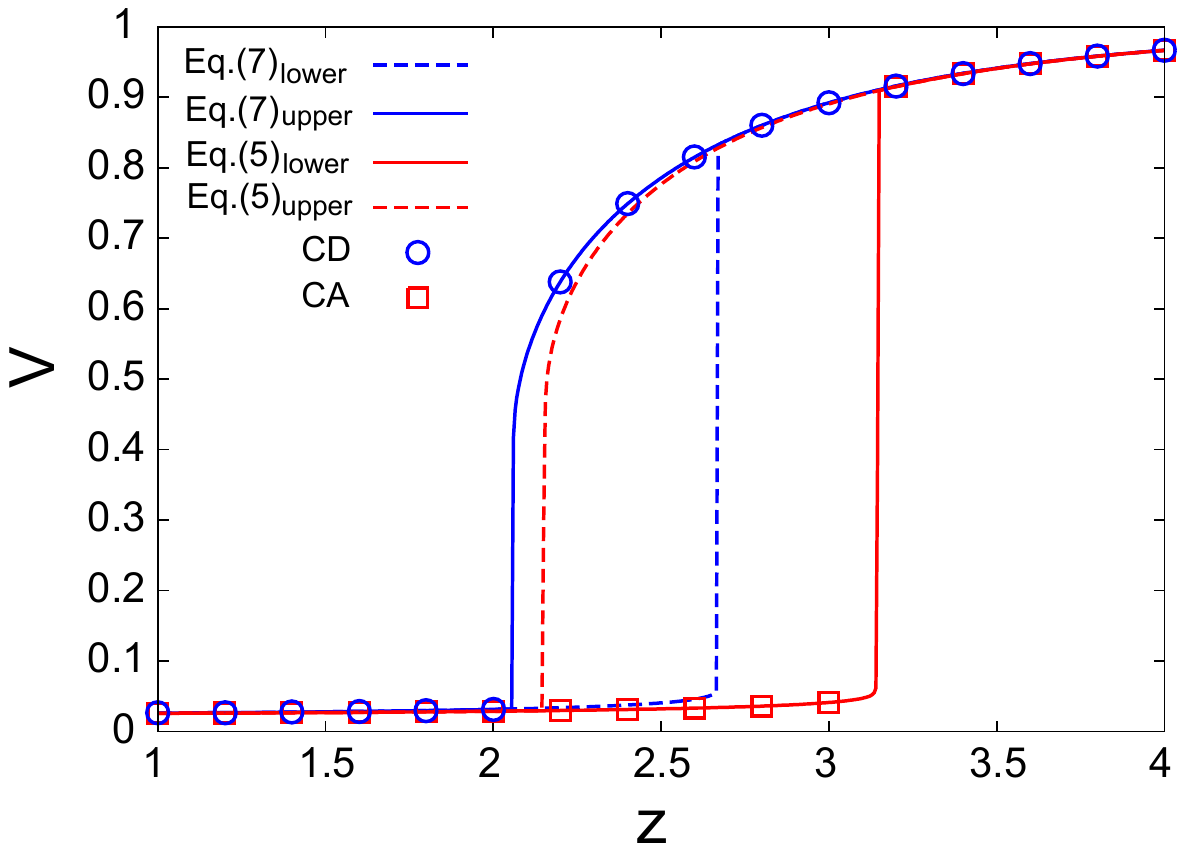}
\caption{(color online)
Solution structure of Eqs.~(4--5) (red lines) and Eqs.~(6--7) (blue lines). Following the definitions of algorithms, the numerical simulation results with $N=10^{6}$ for CA (red square) and CD (blue circle) are presented. Here the each layer of the network is an ER graph with $z_{o}=0.2$ and $\rho=0.02$.}
\end{center}
\label{fig:solution}
\end{figure}

\subsection{Analytic solution for CD algorithm with link overlap}
As noted before, overlapping links play a more intricate role in the CD algorithm.
Since an overlap cluster deactivates as a whole, they behave effectively as a single ``supernode.'' 
Then the condition for a node to be viable becomes whether the supernode to which it belongs 
is connected with other viable supernodes through each and every type of (non-overlapping) links attached to some nodes within the supernode.
To take this into account, it is convenient to ``renomalize'' the network by regarding each overlap cluster as a single supernode,\footnote{Nodes without an overlapping link become isolated supernodes of mass $1$.} which are connected via non-overlapping links connecting nodes in different supernodes. 
This approach was first proposed for the percolation of interdependent networks with link overlap in Ref.~\cite{Hu2013}.

To proceed with computing the viability for the CD algorithm with link overlap
using the renormalized network one further needs two additional quantities. 
First is the distribution $R(m)$ of the mass $m$ (number of nodes) of a supernode by following a randomly chosen non-overlapping link; 
and the second is the distribution $p_m(\vec{q})$ of (non-overlapping) degrees of a supernode of mass $m$ in the renormalized network, where $\vec{q}=(q_1, q_2)$ denotes the (non-overlapping) degrees of a supernode in the two layers. 

Once $R(m)$ and $p_m(\vec{q})$ are at hand, the viability can be obtained as follows: 
The probability that a mass-$m$ supernode including a randomly chosen node does not contain resource nodes is $(1-\rho)^m$.  
Thus the probabilities $u_i$ that the node reached by following  a randomly chosen type-$i$ (non-overlapping) link is not viable satisfy 
\begin{align}
1-u_i=R_\infty &+\sum_{m=1}^{\infty} R(m) \Big\{ [1-(1-\rho)^m] \nonumber\\
&+(1-\rho)^m \sum_{\vec{q}} \frac{q_i p_m(\vec{q})}{z_i} (1-u_i^{q_i-1}) (1-u_j^{q_j}) \Big\}.
\end{align}
Here $R_\infty$ is the probability that the node belongs to the infinite mass supernode (containing the resource node with probability one). 
The first term in the summation gives the probability that the mass-$m$ supernode contains the resource node so that it is viable by itself;
the second term gives the probability that it does not contain the resource node but is connected to other viable nodes (belonging to different supernodes) through each and every layer of non-overlapping links. 
Similarly, the final fraction of viable nodes $V$ can be obtained as 
\begin{align}
V= R_\infty+\sum_{m=1}^{\infty} & R(m) \Big\{ [1-(1-\rho)^m] \nonumber\\
&+(1-\rho)^m \sum_{\vec{q}} p_m(\vec{q}) \prod_{i=1}^2 (1-u_i^{q_i}) \Big\}. 
\end{align}
Note that when $z_o=0$ only $m=1$ term contributes in Eqs.~(6--7), thereby Eqs.~(6--7) and Eqs.~(4--5) become identical.

\begin{figure}[t]
\begin{center}
\includegraphics[width=.95\linewidth]{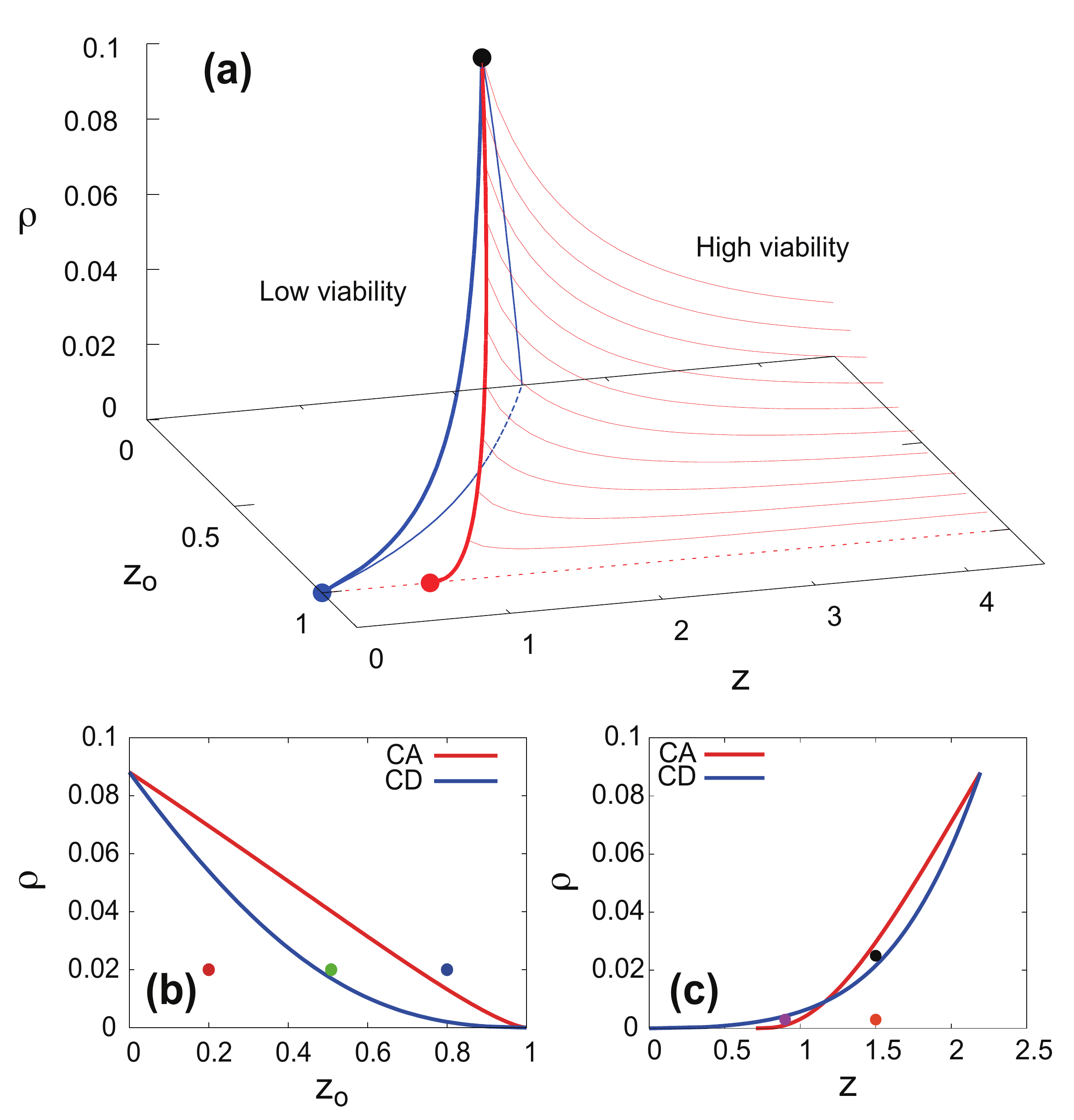}
\caption{
(a) $(\rho, z, z_o)$-phase diagram of viability with link overlap. 
(b) Projections of the phase boundary surface edges (thick lines) for the CA (red) and the CD (blue) on $(\rho, z_o)$ plane.
(c) Same projection plots on the $(\rho, z)$ plane.
}
\end{center}
\end{figure}

The auxiliary quantities $R(m)$ and $p_m(\vec{q})$ can be obtained as follows.
First, assuming that the overlapping and non-overlapping degrees of a node are uncorrelated, the supernode mass distribution $R(m)$ can be obtained from 
the component size distribution  \cite{Newman2007} of the (original) network of overlapping links using the marginal distribution $p_o(k_o)$ for overlapping degrees from $p(k_1,k_2,k_o)$. 
The possibility of non-overlapping links connecting nodes within the same supernode (self-links) and those connecting nodes between two supernodes more than once (multi-links) complicates the computation of $p_m(\vec{q})$.
Fortunately, however, there exist some simplifying conditions. 
One can show that unless the renormalized network has diverging second moments in the supernode degree distribution both can be neglected. 
Under such circumstances and with additional assumption of uncorrelated network layers,
one can obtain $p_m(q_1, q_2)$ by the $m$-th order convolutions of the marginal distribution $p_{12}(k_1,k_2)$ from $p(k_1,k_2,k_o)$

As in the case of Eqs.~(4--5), Eqs.~(6--7) can have two stable solutions. Between the two branches, the outcome of CD algorithm corresponds to the upper branch solution for $V$ of Eqs.~(6--7) (Fig.~4).  
The lower branch solution corresponds to the activation process in terms of the supernode, rather than that of the individual nodes as in the CA algorithm.
Finally, we note that Eqs.~(6--7) for $\rho=0$ are equivalent to the formulation for the mutual percolation with link overlap proposed in Ref.~\cite{Hu2013}, completed with the additional term $R_{\infty}$ which was missing in that study.

\subsection{Application to duplex random graphs}
For explicit illustration, we again consider the duplex random graphs.
Each layer of the network is an ER graph with the mean degrees $z$ and $z_o$ for non-overlapping and overlapping links, respectively.
One may construct such a network by randomly placing $zN$ non-overlapping links on each layer of $N$ nodes,
after which additional $z_oN$ random overlapping links are  placed simultaneously in both layers. Overlaps between non-overlapping and overlapping links can be neglected in large networks.

\begin{figure}[t]
\begin{center}
\includegraphics[width=.9\linewidth]{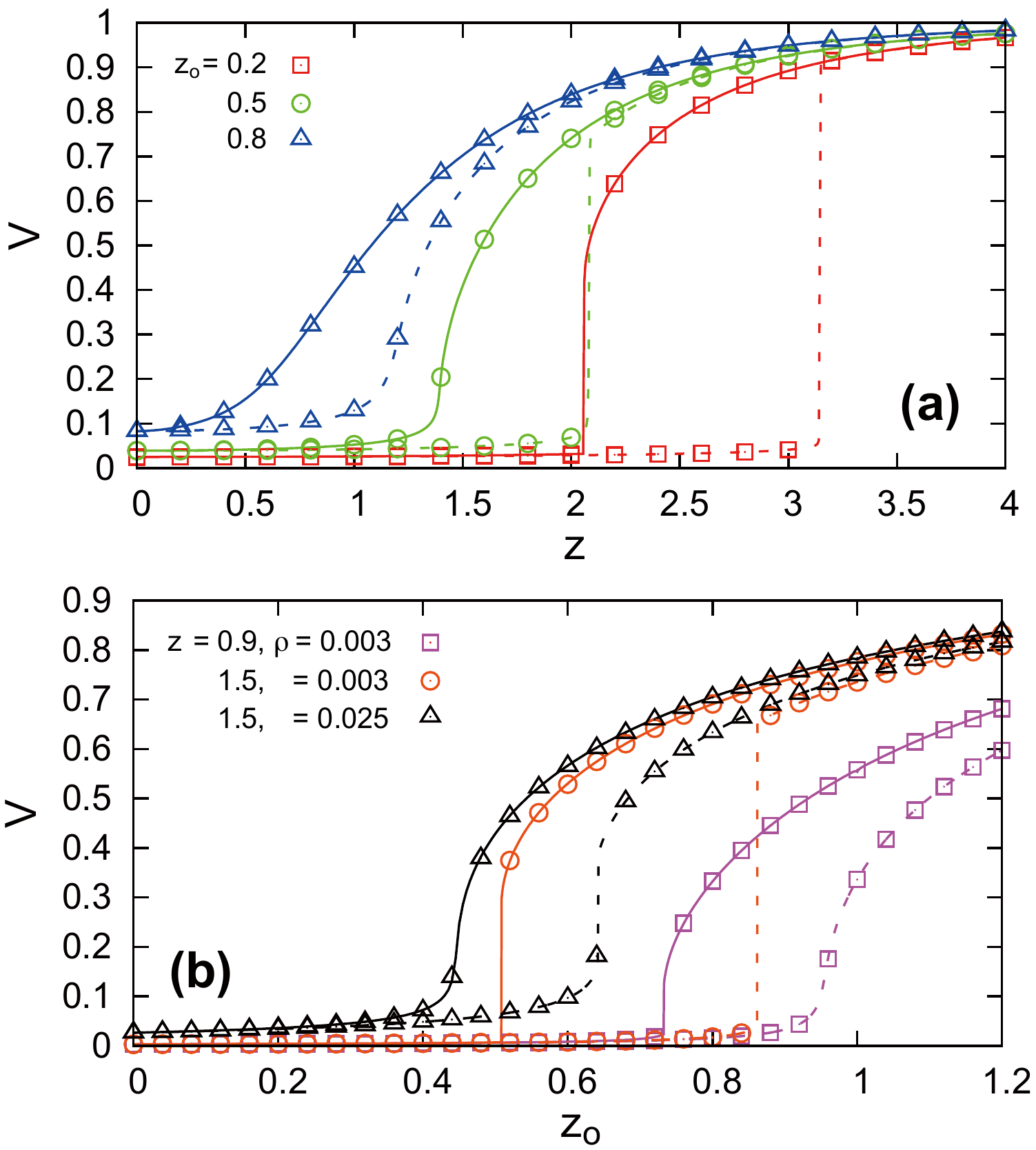}
\caption{
Various hysteretic behaviors of viability with link overlap. 
(a) Depending on the overlapping link mean degree $z_o$ ($\rho=0.02$), combinations of hysteretic behaviors with discontinuous (red), continuous (blue), and mixed (green) changes are observed as the non-overlapping link mean degree $z$ is varied. The parameters and color codes follow those indicated as three dots in Fig.~5(b).
(b) More examples but with varying the overlapping link mean degree $z_o$, combinations of hysteretic behaviors with discontinuous (red) and mixed (black and magenta) changes are observed. The parameters and color codes follow those indicated as three dots in Fig.~5(c).
}
\end{center}
\end{figure}

Eqs.~(4--5) for the viability in the CA algorithm can be written straightforwardly as
\begin{align}
V=\rho+(1-\rho) \Big[(1-e^{-z_o V})+e^{-z_o V} (1-e^{-z V})^2 \Big]. 
\end{align}
Solution for the viability in the CD algorithm can also be obtained by noting that duplex ER graphs satisfy the simplifying conditions for $R(m)$ and $p_m(\vec{q})$ discussed in the previous subsection (with possible exception at $z_o=1$).
Eqs.~(6--7) becomes 
\begin{align}
V=  R_\infty+ \sum_{m=1}^{\infty}  R(m) \Big\{ & [1-(1-\rho)^m]\nonumber\\
&+(1-\rho)^m (1-e^{-zmV})^2 \Big\}, 
\end{align}
where $R(m)=(m z_o)^{m-1} e^{-m z_o}/m!$ and $R_{\infty}=1-\sum_{m=1}^{\infty}R(m)=1+W(-z_oe^{-z_o})/z_o$ with $W(\cdot)$ being Lambert $W$-function \cite{NewmanBook}.

Using the lower branch solution of Eq.~(8) and the upper branch solution of Eq.~(9), 
one can construct the three dimensional phase diagram in the $(\rho, z, z_o)$ parameter space (Fig.~5). 
The phase diagram is rather complicated, which should read as follows:
The system is in high (low) viability phase on the right (left) side of red (blue) surface. 
The two phase boundary surfaces correspond to the loci at which lower branch solution loses stability in Eq.~(8) and discontinuous jump of viability in CA algorithm occurs (red) and those at which upper branch solution in Eq.~(9) and CD algorithm do so (blue). 
As the parameters are varied, discontinuous changes in the viability occur when the parameter trajectory penetrates the phase boundary surface.
On the other hand, if the parameter trajectory does not penetrate the boundary surface, the viability changes smoothly. 
Therefore one can observe diverse sets of hysteretic behaviors, depending on the way the system parameters are varied (Fig.~6). The facilitating effect of the overlapping links for viability can also be seen from the phase diagram. 

\section{Mutual percolation with link overlap}
\label{sec:mutual}
As noted, mutual percolation corresponds formally to the viability problem with $\rho=0$. 
Let us look at this correspondence more closely. Without link overlap, one can obtain the equations for mutual percolation \cite{Son2012} by simply setting $\rho=0$ in Eqs.~(1--2). 
For $\rho=0$, the lower branch solution of Eqs.~(1--2) becomes a trivial one $V=0$. 
Above $z_{\text{CD}}(\rho=0)=z_{\text{MP}}$, where $z_{\text{MP}}$ denotes the mutual percolation transition point without link overlap, the nontrivial solution appears in the upper branch. 
Therefore the size of giant mutual component $M$ in mutual percolation can be identified with the upper branch solution for $V$ of Eqs.~(1--2) with $\rho=0$.

In the presence of link overlap, one should take into account the distinctive role of the overlapping links forming the mutual component by themselves. 
This can only be implemented by the CD-type Eqs.~(6--7), and cannot be properly accounted for with the CA-type Eqs.~(4--5). Note that the upper branches of the two approaches do not converge when $z_o>0$.
Hence we obtain the analytic solution for the mutual percolation with link overlap as follows.
The probability $v_i$ that a node reached by following a randomly chosen link in $i$-layer
does not belong to the giant mutual component is given by setting $\rho= 0$ of Eq.~(6) as  
\begin{align}
1-v_{i}&=R_\infty +\sum_{m=1}^{\infty}R(m) \sum_{\vec{q}} \frac{q_i p_m(\vec{q})}{\langle q_i\rangle} (1-v_i^{q_i-1})(1-v_j^{q_j}),
\end{align}
using the same auxiliary quantities $R(m)$ and $p_m(\vec{q})$.
Then the size of the giant mutual component $M$ is obtained in terms of the probability that a randomly chosen node belongs to the giant mutual component by 
\begin{align}
M&=R_\infty+ \sum_{m=1}^{\infty} R(m)  \sum_{\vec{q}} p_m(\vec{q}) \prod_{i=1}^2 (1-v_i^{q_i}),
\end{align}
which is the $\rho=0$ case of Eq.~(7) with $M$ and $v_i$'s in place of $V$ and $u_i$'s, respectively. We note once again that these Eqs.~(10--11) are equivalent to the formulation proposed in Ref.~\cite{Hu2013}, except for the additional term $R_{\infty}$ which was missing in that study.

\begin{figure}[t]
\begin{center}
\includegraphics[width=.85\linewidth]{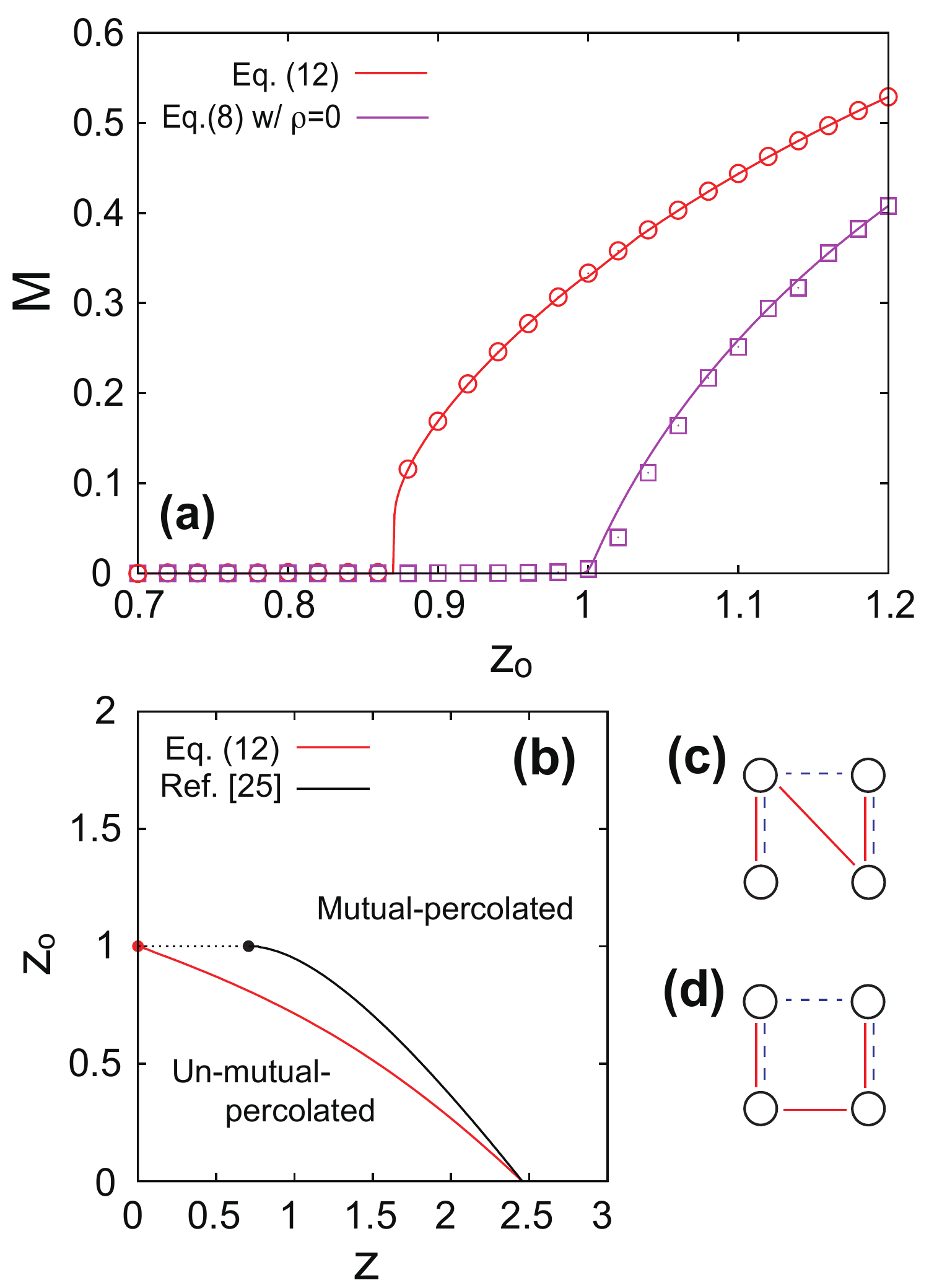}
\caption{
(a) Mutual percolation of the duplex ER graphs as a function of the overlapping link mean degree $z_o$ for fixed $z=0.5$ (red). Symbols (circles) are numerical results with $N=10^{6}$. Comparison with the predictions from Eq.~(8) (line) and the message-passing algorithm proposed in Ref.~\cite{Cellai2013} (squares) is made (purple). 
%{\bf [NEED simulation results for $N=10^6$]}
(b) The $(z, z_o)$ phase diagram for the mutual percolation of duplex ER graphs. Phase boundary obtained from Eq.~(12) (red) is displayed, together with that from Ref.~\cite{Cellai2013} (black lines), shown for comparison. Solid lines represent lines of discontinuous transitions which terminate at the filled circle. Dotted line is the continuous transition line proposed in \cite{Cellai2013}.
(c, d) Two example configurations of mutual-percolated clusters. }
\end{center}
\end{figure}

\subsection{Application to duplex random graphs}
Applying Eqs.~(10--11) to duplex random graphs with non-overlapping link mean degree $z$ and overlapping link mean degree $z_o$, one  obtains from Eq.~(9) the following equation for the giant mutual component size $M$,
\begin{equation}
M=R_\infty+\sum_{m=1}^{\infty} R(m)(1-e^{-m z  M})^2,
\end{equation}
with $R_{\infty}$ and $R(m)$ given in Sec.~3.3. 

Solving Eq.~(12) for $M$, one can analyze the mutual percolation behaviors, showing good agreement with numerical simulations [Fig.~7(a)].
One can also obtain the $(z, z_o)$-phase diagram for the mutual percolation [Fig.~7(b)], which
predicts discontinuous mutual percolation transitions except at $(z, z_o)=(0, 1)$, which is nothing but the continuous ordinary percolation transition point. 
This prediction is in agreement with the claims in \cite{Hu2013} but in contrast with those of \cite{Cellai2013}, which predicted the existence of tricritical point [Fig.~7(b)].
As noted in Sec.~3.1, the approach of \cite{Cellai2013} corresponds to the CA-type Eqs.~(4--5) with $\rho=0$ (upper branch), which has different solution than that of Eqs.~(10--11) unless $z_o=0$ [Fig.~7(a)]. Being based on local update, this approach is able to detect mutual connectivity in configurations like Fig.~7(c), but can fail for configurations like Fig.~7(d).  

\begin{figure}[t]
\begin{center}
\includegraphics[width=.85\linewidth]{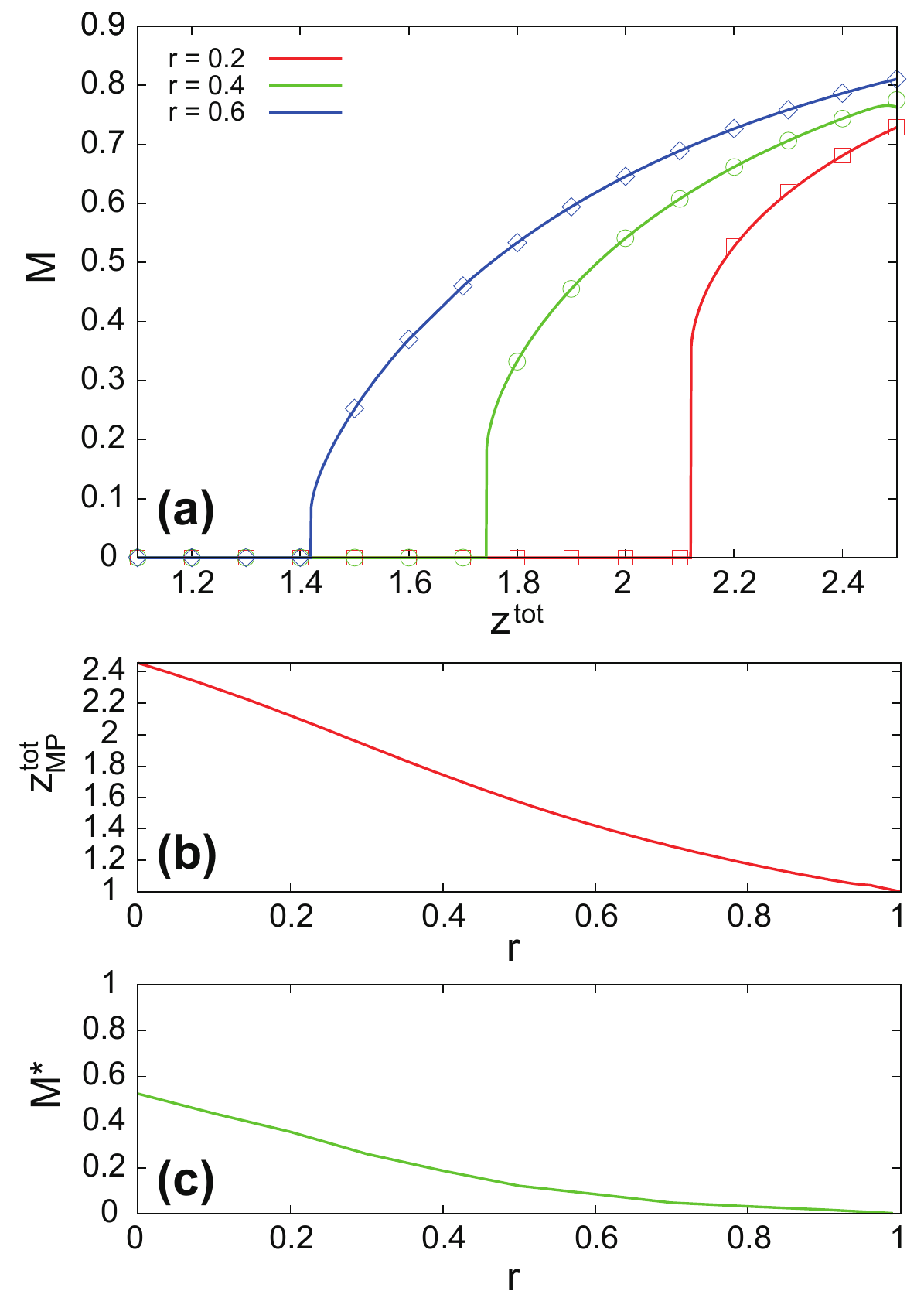}
\caption{
(a) Mutual percolation of duplex ER graphs as a function of the total mean degree of each layer $z^{tot}=z+z_o$ for various overlap fractions $r=0.2, 0.4, 0.6$ (from right to left). Lines are obtained by Eq.~(12) and symbols are numerical results with $N=10^{6}$.
(b, c) Plots of the total mean degree for mutual percolation transition $z^{tot}_{\text{MP}}$ (b) and the size of discontinuous jump in the giant mutual component size $M^*$ at the transitions (c) as functions of the overlap fraction $r=z_o/z^{tot}$.
}
\end{center}
\end{figure}

Another view at the facilitating role of link overlap for mutual connectivity is provided by  examining how the transition points for mutual percolation of the total mean degree $z_{\text{MP}}^{tot}$, where $z^{tot}= z+z_o$, changes with the overlap fraction $r\equiv z_o/z^{tot}$ (Fig.~8). For the same total mean degree $z^{tot}$, the giant mutual component size increases with the overlap fraction $r$ [Fig.~8(a)]. As the overlap fraction $r$ increases, the mutual percolation transition occurs at lower $z_{\text{MP}}^{tot}$, from $z_{\text{MP}}^{tot}(r=0)=z_{\text{MP}}=2.455407\cdots$ to $z_{\text{MP}}^{tot}(r=1)=1$ [Fig.~8(b)]. The size of discontinuous jump in the giant mutual component size at the transition decreases with the overlap fraction, from $M^*(r=0)=0.511699\cdots$ to $M^*(r=1)=0$ (that is, continuous) [Fig.~8(c)].

\section{Implications to robustness against link failures}
\label{sec:robustness}

Given the distinctive role of overlapping links for multiplex viability, it is of interest to examine how differently the network would respond to failures of the two classes of links.
Having established the analytic approaches for the viability and mutual percolation in previous sections, one can readily employ the existing method for examining the effect of random link failures \cite{Callaway,Min2014a}.

Here we will present analytic formulae for the case of mutual percolation as they are similar to yet simpler than those for the viability.
Suppose $1-f_o$ fraction of overlap links is removed, the degree distribution of overlapping links 
becomes $\tilde{p}_o(k')=\sum_{{k_o}={k'}}^{\infty} p_o(k_o) \binom {k_o} {k'} (f_o)^{k'} (1-f_o)^{k_o-k'}$,
and consequently the mass distribution of supernodes changes accordingly.
Then the giant mutual component size can be obtained by solving Eqs.~(10--11) with the supernode mass distribution $\tilde{R}(m)$ corresponding to $\tilde{p}_o(k')$. 
On the other hand, when the non-overlapping links are removed, 
Eqs.~(10--11) for $v_i$ and $M$ should be modified slightly, while $R(m)$ remains unchanged.
Suppose $1-f_i$ fraction of $i$-type non-overlapping links is removed, 
the probability $v_i$ that a node reached by following 
a randomly chosen non-overlapping link in $i$-layer does not belong to the giant mutual component becomes \cite{Min2014a}
\begin{align}
1-v_{i}&=R_\infty +\sum_{m=1}^{\infty}R(m) \sum_{\vec{q}} \frac{q_i p_m(\vec{q})}{\langle q_i\rangle}\nonumber\\ 
&\{ 1-[1+(v_i-1)f_i]^{q_i-1} \}  \{ 1-[1+(v_j-1)f_j]^{q_j} \},
\end{align}
for $i=1,2$.
Then the giant mutual component size is obtained by 
\begin{align}
M=R_\infty +\sum_{m=1}^{\infty}R(m) \sum_{\vec{q}} p_m(\vec{q}) \prod_{i=1}^2\{ 1-[1+(v_i-1)f_i]^{q_i} \} .
%&\times\{ 1-[1+(v_j-1)f_j]^{q_j} \}.
\end{align}

\begin{figure*}
\begin{center}
\includegraphics[width=0.85\linewidth]{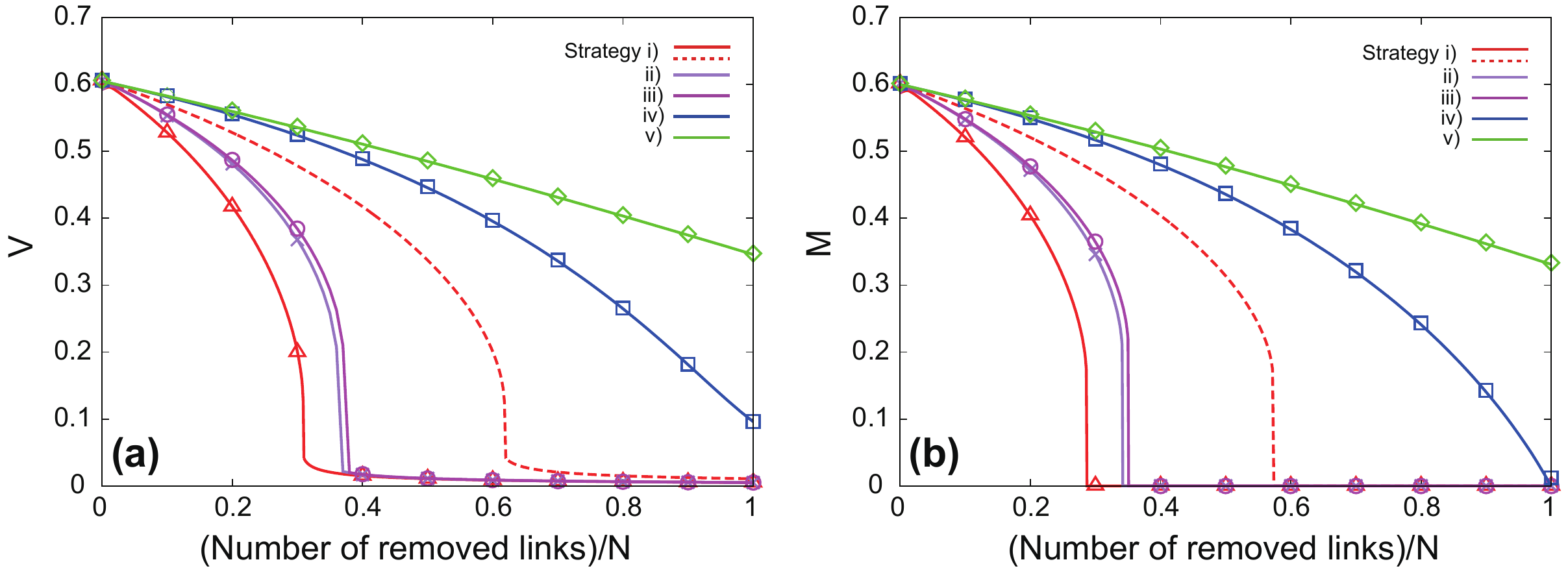}
\caption{
The viability (a) and the giant mutual component size (b) of duplex ER graphs of unperturbed mean degrees $z_1=z_2=z_o=1.0$ as a function of the number of removed links for different removal strategies described in the text. The fraction of resource nodes in (a) is $\rho=0.005$. Lines are obtained from Eqs.~(15--17) and symbols are from simulations with $N=10^6$ averaged over $10^2$ runs. Plots for the strategy i) are drawn twice, one by counting the overlapping link as one link (solid) and the other by counting it as two links (dotted).}
\end{center}
\end{figure*}

\subsection{Application to duplex random graphs}
For ER layers, the supernode mass distribution $R'(m)$ after the random removal of $1-f_o$ fraction of overlapping links of unperturbed mean degree $z_o$ is given by 
\begin{align}
\tilde{R}(m)=\frac{(m z_{o}f_o)^{m-1}\cdot e^{-m z_{o}f_o}}{m!}.
\end{align}
Then the giant mutual component size is obtained by
\begin{align}
M &=\tilde{R}_{\infty}+ \sum_{m=1}^{\infty}\tilde{R}(m)(1-e^{-z_1 m M})(1-e^{-z_2 m M}),
\end{align}
with $\tilde{R}_{\infty}=1-\sum_{m=1}^{\infty}\tilde{R}(m)$ and $\tilde{R}(m)$ from Eq.~(15).
For the random removal of non-overlapping links, Eqs.~(13--14) reduces into a single equation for the giant mutual component size as
\begin{align}
M &=R_{\infty} +\sum_{m=1}^{\infty}R(m)(1-e^{-z_{1}f_1 m M})(1-e^{- z_{2}f_2 mM}).
\end{align}

We consider a variety of random link removal strategies: i) Removing the overlapping link simultaneously; ii) removing only the type-1 link among the overlapping link; iii) removing any one link among the overlapping link; iv) removing only the type-1 non-overlapping link; v) removing overlapping link of any type (Fig.~9).
We found as expected that overlapping link removals are significantly more detrimental than non-overlapping link removal in both system's viability [Fig.~9(a)] and mutual connectivity [Fig.~9(b)]. 
Surprisingly, removals of only one of the overlapping links are quite closely as efficient as the removals of the entire overlapping links, reflecting the critical role of the link overlap.
Such a critical role of link overlap may provide mechanistic basis for understanding of the importance of multiplexity in social context \cite{Minor1983}.
Finally, it is also observed that removing the non-overlapping links in a particular layer is more efficient to break the network down than the removals of the same number of non-overlapping links from any layers.

\section{Summary and discussions}
\label{sec:conclusion}

In this article, we have studied the effect of link overlap in the viability of multiplex networks, both analytically and numerically. Distinctive role of overlapping links in viability and mutual connectivity is emphasized and exploited for setting up analytic framework properly.
Especially, it was shown that with link overlap the CA and the CD algorithms for the viability are described by different equations. 
As a result, a rich phase diagram for viability is obtained and the hysteretic patterns of viability become greatly diversified in the presence of link overlap.
Mutual percolation with link overlap is revisited as the $\rho=0$ limit of the CD case of multiplex viability problem, and controversy between existing results was clarified. For duplex random graphs,  the transition is obtained to be  discontinuous as long as $z>0$ at the transition, thus the tricritical point is absent. 
The distinctive role of overlapping links was demonstrated explicitly by the different responses of the network under random removals of overlapping and non-overlapping links, respectively, as well as under several removal strategies.

Our results show that the link overlap strongly facilitate viability and mutual percolation as the overlapping links form mutually-connected cluster by themselves. At the same time, by forming cluster the effect of link overlap can become long-ranged, posing challenge in analytical approach to the problem. The analytical approach for CD algorithm based on Eqs.~(6--7) and Eqs.~(10--11) can be extended easily for general $n$-layers in case of the complete link overlap, {\it i.e.}, when the overlapping links span across all the layers.
In case of the partial overlap, for instance when some links overlap across only two of three layers, the present approach cannot be readily applied, although the numerical algorithm can. 
It is left an open problem to establish theory for the CD algorithm with general overlaps. The theory for CA algorithm, Eqs.~(4--5), however, would be extended to $n$-layers straightforwardly by generalizing the degree vector to include the partially overlapping link degrees as well.

Speaking in clich\'e, the link overlap is more than sum of its part links. 
The impact of link overlap on multiplex network dynamics is expected be seen widespread given its strong empirical evidences across diverse real-world systems and generic nontrivial role for cooperative couplings between layers. These may include, but not limited to, cascading failure \cite{Sandpile}, cooperative infection \cite{Chen2013}, information spreading \cite{Min2013}, and social behavioral cascade \cite{KMLee2014}, opening a wide opportunity for future studies. Lastly, the CD and CA algorithms for the multiplex viability can be thought of as multiplex generalizations of the classic bootstrap percolation \cite{Chalupa79} and diffusion percolation \cite{Adler88}, respectively. In this respect, the study of viability problem in low dimensions could be of interest, too.

\section{Acknowledgements}
Discussions with Davide Cellai and Ginestra Bianconi on the issues discussed in Sec.~4 are warmly acknowledged. We thank G.\ Bianconi also for sharing the message-passing algorithm code used in Fig.~7(a). This work was supported by the Basic Science Research Program through an NRF grant funded by MSIP (Grant No. 2011-0014191).

\end{document}